\documentclass{aa}
\usepackage{amssymb,amsmath}
\usepackage{graphicx}
\usepackage[varg]{txfonts}
\usepackage[colorlinks=true, citecolor=blue, linkcolor=blue, urlcolor=blue]{hyperref}
\usepackage{natbib}
\bibpunct{(}{)}{;}{a}{}{,} 

\begin{document} 

\title{The strongest gravitational lenses: IV. The order statistics of the largest Einstein radii with cluster mergers}

\titlerunning{Order statistics of the largest Einstein radii with cluster mergers}

\author{
M. Redlich \inst{1}
\and
J.-C. Waizmann \inst{2}
\and
M. Bartelmann \inst{1}
}

\authorrunning{M. Redlich et al.}

\institute{
Zentrum f\"ur Astronomie der Universit\"at Heidelberg, Institut f\"ur 
Theoretische Astrophysik, Albert-Ueberle-Str.~2, 69120 Heidelberg, Germany \\
\email{matthias.redlich@stud.uni-heidelberg.de}
\and
Blue Yonder GmbH, Ohiostrasse 8, D- 76149 Karlsruhe, Germany
}

\date{\emph{A\&A manuscript, version \today}}

\abstract{Based on techniques developed in the previous papers of this series, we investigate the impact of galaxy-cluster mergers on the order statistics of the largest Einstein radii. We show that the inclusion of mergers significantly shifts the extreme value distribution of the largest Einstein radius to higher values, typically increasing the expected value by ${\sim}10\%$. A comparison with current data reveals that the largest observed Einstein radius agrees excellently well with the theoretical predictions of the $\Lambda$CDM model at redshifts $z > 0.5$. At redshifts $z < 0.5$, our results are somewhat more controversial. Although cluster mergers also increase the expected values of the order statistics of the $n$ largest Einstein radii by ${\sim}10\%$, the theoretically expected values are notably lower (${\sim}3\sigma$ deviation for $n = 12$) than the largest Einstein radii of a selected sample of SDSS clusters in the redshift range $0.1 \leq z \leq 0.55$. The uncertainties of the observed Einstein radii are still large, however, and thus the measurements need to be carefully revised in future works. Therefore, given the premature state of current observational data, overall, there is still no reliable statistical evidence for observed Einstein radii to exceed the theoretical expectations of the standard cosmological model.}

\keywords{Cosmology: theory -- Gravitational lensing: strong -- Galaxies: clusters:general --  Methods: statistical}

\maketitle

\section{Introduction}
\label{sec:sl4:introduction}

Do the strongest observed gravitational lenses exceed the theoretical expectations of the standard cosmological model? This question has long been debated in the literature \citep[see][for reviews]{2010CQGra..27w3001B, 2013SSRv..177...31M}, and was the central theme of this series of papers.

In our first work \citep[][hereafter paper I]{2012A&A...547A..66R}, we developed a new semi-analytic method for populating a fictitious observer's past null cone with merging galaxy clusters. We studied the impact of cluster mergers on the optical depth for giant gravitational arcs and the distribution of Einstein radii, finding that mergers are particularly relevant for the strongest gravitational lenses in the Universe. Furthermore, we showed that strong-lensing cross sections for giant gravitational arcs are tightly correlated with Einstein radii. This correlation allowed us to infer optical depths from distributions of Einstein radii, and also justified why we focused on the statistics of Einstein radii in the subsequent works. 

In \citet{2012A&A...547A..67W} (hereafter paper II), we developed a framework for applying the theory of extreme value statistics to the distribution of the single largest Einstein radius in a given cosmological volume. We found that the extreme value distribution is particularly sensitive to the precise choice of the halo mass function, lens triaxiality, the inner slope of the halo density profile, and the mass-concentration relation. Most importantly, and in contrast to many previous studies, we demonstrated that the largest known Einstein radius -- which was observed in the galaxy cluster MACS J0717.5+3745 \citep{2009ApJ...707L.102Z, 2011MNRAS.410.1939Z, 2012A&A...544A..71L, Medezinski2013} -- is consistent with the theoretical expectations of the standard cosmological model.

Given the insight that the occurrence probability of the single largest Einstein radius is strongly influenced by several theoretical uncertainties, we argued that it would be desirable to formulate $\Lambda$CDM exclusion criteria that are based on a number of observations instead of a single event. Therefore, we extended our theoretical framework to higher order statistics in the third work \citep[][hereafter paper III]{2014A&A...565A..28W}. More precisely, we studied exclusion criteria based on the individual and joint order distributions of the $n$ largest Einstein radii. Most importantly, again, our computations revealed that the large Einstein radii of the twelve high-redshift ($z > 0.5$) MACS clusters analysed by \citet{2011MNRAS.410.1939Z} do not exceed the expectations of the $\Lambda$CDM model.

However, so far, our analysis is still incomplete in the following sense: while we explicitly emphasised the importance of cluster mergers for the statistics of the strongest gravitational lenses in paper I, we did not include mergers in our computations for papers II and III. This decision was motivated by several factors. Firstly, particularly in paper II, we investigated the properties of triaxial gravitational lenses (such as orientation and profile shape) that decisively affect the extreme value distribution of the largest Einstein radius. For these studies, it was important to isolate individual effects, and therefore it was neither desirable nor required to include cluster mergers. Secondly, we aimed to derive conservative exclusion criteria. We were able to show that the strongest observed gravitational lenses do not exceed the theoretical expectations of the standard cosmological model even if cluster mergers are neglected. Because mergers additionally boost the strong lensing efficiency, including these events should additionally consolidate this conclusion. Thirdly, including clusters mergers is computationally very expensive. A typical realisation of a mock universe requires Monte Carlo simulations of $\sim 10^6$ merger trees. In addition, lensing computations with multiple haloes in the field of view are substantially more expensive than those with single haloes. To estimate the order statistics of the largest Einstein radii with an acceptable precision, $\sim 10^3$ mock universes have to be sampled. This whole procedure is computationally very demanding, even though our semi-analytic method was specifically tailored to be fast.

In this sense, this paper completes the present series. Based on previously developed techniques, we compute the impact of cluster mergers on the order statistics of the largest expected Einstein radii in the Universe. This is not only interesting in itself, but also because it differs from the analysis performed in paper I, where we investigated the impact of cluster mergers on averaged quantities such as the optical depth for giant gravitational arcs. Moreover, we compare our theoretical results to another, independent set of observational data at redshifts $z < 0.5$ \citep{2012MNRAS.423.2308Z}, finding that the observed gravitational lenses in this cosmological volume might be somewhat stronger than theoretically expected.

The plan for this paper is as follows: In Sect.~\ref{sec:sl4:theoretical_basis}, we briefly summarise the most important theoretical concepts that were introduced in our previous works. Section~\ref{sec:sl4:merger_algorithm} describes the merger tree algorithm used for this work, and also contains some information on speeding up the computations. The impact of cluster mergers on the order statistics of the largest Einstein radii is analysed in Sect.~\ref{sec:sl4:extreme_order_statistics_mergers}. Thereafter, in Sect.~\ref{sec:sl4:comparison_observations}, we compare our theoretical predictions with current observational data. In Sect.~\ref{sec:sl4:conclusions}, we summarise our main results and finally conclude. 

Throughout this work, we adopt the best-fitting cosmological parameters $(\Omega_{\Lambda 0}, \Omega_{{\mathrm{m}}0}, \Omega_{{\mathrm{b}}0}, h, 
\sigma_8) = (0.685, 0.273, 0.047, 0.673, 0.829)$, which were derived from the Planck 2013 data \citep{2013arXiv1303.5076P}.

\section{Theoretical basis}
\label{sec:sl4:theoretical_basis}

For convenience, we briefly summarise the most important theoretical concepts this work is based upon. We refer to our previous publications for more thorough introductions to triaxial gravitational lenses (paper I) as well as extreme value (paper II) and order statistics (paper III).

\subsection{Triaxial gravitational lenses}
\label{subsec:sl4:triaxial_lenses}

As extensively discussed in paper I, the triaxiality of galaxy clusters is very important for their strong-lensing properties. We therefore adopt the triaxial, generalised NFW profile proposed by \citet{2002ApJ...574..538J} to model the dark matter density profile of galaxy clusters:
\begin{align}
\label{eq:sl4:triaxial_density_profile}
\rho(R) &= \frac{\delta_{\mathrm{ce}} \; \rho_{\mathrm{crit}}(z)}{(R/R_0)^{\alpha}(1+R/R_0)^{3-\alpha}} \; , \\ 
\label{eq:sl4:elliptical_radius}
R^2 &\equiv \frac{{x'}^2}{(a/c)^2} + \frac{{y'}^2}{(b/c)^2} + {z'}^2 \qquad \left( a \leq b \leq c\right) \; .
\end{align}
Here, $z$ denotes the halo redshift, $R_0$ is a scale radius, $\delta_{\mathrm{ce}}$ is a characteristic density, and $\rho_{\mathrm{crit}}(z)$ denotes the critical density of the universe; see paper I for more detail on these quantities, the sampling of random density profiles, and the computation of the lensing signal.

In this work, we set the inner slope $\alpha$ of the density profile to the standard NFW value $\alpha = 1$. As detailed in Sect. 5.2 of paper II, we force the scaled axis ratios $a_{\mathrm{sc}}$ (minor axis) to lie within the $99\%$ confidence interval of the probability density function (PDF) for $a_{\mathrm{sc}}$ derived by \citet{2002ApJ...574..538J}. By doing so, we avoid unrealistic density profiles with extremely low axis ratios and too low concentrations.

\subsection{Definition of the Einstein radius}
\label{subsec:sl4:einstein_radius}

Axially symmetric gravitational lenses exhibit circular tangential critical curves, whose size can naturally be quantified by means of their Einstein radius. In contrast, the tangential critical curves of merging triaxial gravitational lenses are irregular and cannot be described by simple geometric figures (cf. paper I). Nevertheless, it has been proven useful to generalise the concept of an Einstein radius to arbitrarily shaped critical curves.

In this work, we use a geometrically motivated definition. Let $A$ denote the area enclosed by the tangential critical curve of an arbitrary gravitational lens. The effective Einstein radius $\theta_{\mathrm{eff}}$ of this lens is then defined by
\begin{equation}
\label{eq:sl4:def_theta_eff}
\theta_{\mathrm{eff}} \equiv \sqrt{\frac{A}{\pi}} \; ,
\end{equation}
so that a circle with radius $\theta_{\mathrm{eff}}$ has the same area $A$.

\subsection{Extreme value statistics}
\label{subsec:sl4:extreme_value_statistics}

The theory of extreme value statistics (EVS) describes the stochastic behaviour of extremes and provides a rigorous framework for determining the likelihood of rare events \citep{gumbel1958statistics}. In particular, the theory defines the mathematical recipe for quantitatively answering questions of the following kind: Given a cosmological model, how likely is it that the largest Einstein radius in a certain cosmological volume is as large as $X \arcsec$? 

The method developed in paper II is based on the so-called Gnedenko approach, which models the distribution of block maxima \citep{1928PCPS...24..180F, gnedenko1943limited}. In simple words, it can be summarised as follows: Suppose that we intend to calculate the extreme value distribution of the largest Einstein radius in a certain cosmological volume. To this end, we populate the volume with a mock catalogue of triaxial gravitational lenses and compute their Einstein radii. We note the largest Einstein radius $\theta_{\mathrm{eff}}^1$ of the current realisation, and repeat this random process $n$ times. After this procedure, we have a set of $n$ largest Einstein radii, $M = \left( \theta_{\mathrm{eff}}^1, ..., \theta_{\mathrm{eff}}^n \right)$. As was shown in paper II, for large $n$, the cumulative distribution function (CDF) of these maxima converges to the general extreme value (GEV) distribution
\begin{equation}
\label{eq:sl4:GEV}
  G_{\gamma,\,\beta,\,\alpha}(x) = \left\{ 
  \begin{array}{l l}
    \exp{\left\lbrace -\left[1+\gamma \left(\frac{x-\alpha}{\beta}\right)\right]^
    {-1/\gamma}\right\rbrace}, & \quad {\mathrm{for}}\quad\gamma\neq 0,\\
    \exp{\left\lbrace \mathrm{e}^{-\left(\frac{x-\alpha}{\beta}\right)}\right\rbrace} ,& \quad 
    {\mathrm{for}}\quad\gamma = 0,\\
  \end{array} \right.
\end{equation}
with the location, scale and shape parameters $\alpha$, $\beta$, and $\gamma$.

Fitting the GEV to the empiric CDF has several advantages. Amongst others, the fit allows us to derive analytic relations for probabilities and exclusion criteria, and also enables us to smoothly extrapolate to regions that are only sparsely sampled by the empiric data.

\subsection{Order statistics}
\label{subsec:sl4:order_statistics}

In paper III, we argued that the GEV of the single largest Einstein radius is sensitive to several theoretical uncertainties and that it might therefore be advisable to formulate more robust exclusion criteria based on $n$ observations instead of a single one. To do this, we extended our framework to the order statistics of the $n$ largest Einstein radii in a given cosmological volume.

For a random sample of $n$ Einstein radii  $\theta_{\mathrm{eff}}^1, \theta_{\mathrm{eff}}^2, ..., \theta_{\mathrm{eff}}^n$, the order statistic is given by the ordered set $\theta_{\mathrm{eff}}^{(1)} \leq  \theta_{\mathrm{eff}}^{(2)} \leq ... \leq \theta_{\mathrm{eff}}^{(n)}$, so that $\theta_{\mathrm{eff}}^{(1)}$ and $\theta_{\mathrm{eff}}^{(n)}$ denote, respectively, the smallest (minimum) and the largest (maximum) Einstein radius in the sample. The CDF of the $i$-th order is
\begin{equation}
\label{eq:sl4:F_order}
F_{(i)}\left( \theta_{\mathrm{eff}} \right) = \sum_{k=i}^n\binom{n}{k}\left[F \left( \theta_{\mathrm{eff}} \right) \right]^k\left[1-F \left( \theta_{\mathrm{eff}} \right)\right]^{n-k}\, ,
\end{equation}
where $F \left(\theta_{\mathrm{eff}} \right)$ denotes the CDF the sample of Einstein radii was drawn from. As discussed above (and, more extensively, in paper III), the distribution function of the maxima $F_{(n)}\left( \theta_{\mathrm{eff}} \right)$ converges to the GEV for large sample sizes
\begin{equation}
F_{(n)}\left( \theta_{\mathrm{eff}} \right) = \left[ F\left( \theta_{\mathrm{eff}} \right) \right]^n \to G_{\gamma,\,\beta,\,\alpha}\left( \theta_{\mathrm{eff}} \right) \; ,
\end{equation}
which, in a certain domain, allows us to relate the underlying distribution function $F \left(\theta_{\mathrm{eff}} \right)$ to the GEV and, using Eq.~\eqref{eq:sl4:F_order}, derive analytic expressions for the CDFs of the second largest, third largest, etc. Einstein radius.

\section{Algorithm for including cluster mergers}
\label{sec:sl4:merger_algorithm}

\subsection{Extended Press-Schechter merger tree algorithm}
\label{subsec:sl4:pch_merger_trees}

\begin{figure}
	\centering
	\includegraphics[width=0.45\textwidth]{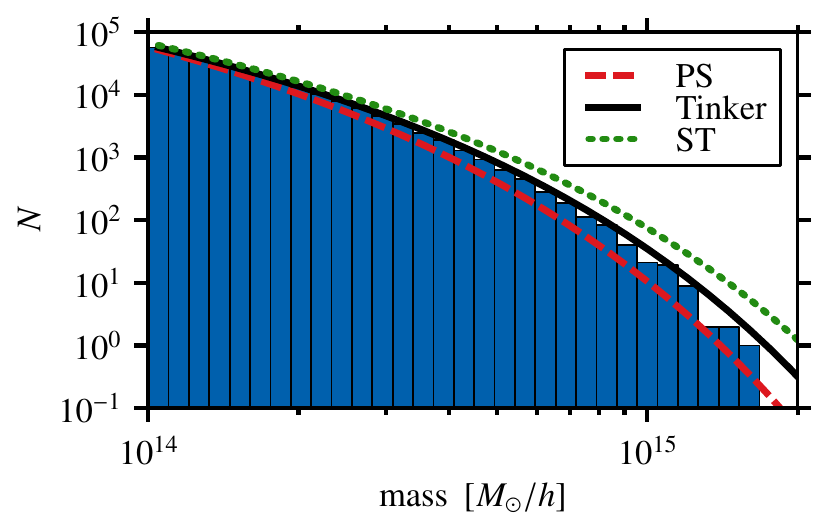}
	\caption[Exemplary mass function generated with the PCH merger tree algorithm.]
	{Comparison between a Monte-Carlo sampled halo catalogue in the redshift range $0.5 < z < 1.0$ (full sky) and the corresponding theoretically expected values of different halo mass functions. The random halo catalogue was generated with the merger tree algorithm of \citet{2008MNRAS.383..557P}. The red dashed, solid black, and green dotted curves indicate the mass functions proposed by \citet{1974ApJ...187..425P} (PS), \cite{TinkerMF} (Tinker), and \citet{ShethTormen1999} (ST).}
	\label{fig:sl4:PCH_mass_function}
\end{figure}

In paper I, we developed a semi-analytic method that allows us to project merging galaxy clusters onto a fictitious observer's past null cone and calculate their strong-lensing properties. Our method can be briefly summarised as follows: First, we specify the redshift range $\left[ z_{\mathrm{min}}, z_{\mathrm{max}} \right]$ and the sky fraction $f_{\mathrm{sky}}$ that we are about to analyse, and compute the size of the comoving volume that corresponds to these parameters. Next, we envisage this cosmological volume at redshift $z = 0$ and populate it with a representative halo population by sampling from a halo mass function. We assign each halo a (properly weighted) random observation redshift $z_{\mathrm{obs}}$ $\left( z_{\mathrm{min}} < z_{\mathrm{obs}} < z_{\mathrm{max}} \right)$ and use a merger tree algorithm to evolve the halo backwards in time up to $z_{\mathrm{obs}}$. A merger tree simulation is a Monte-Carlo process that starts with the initial halo mass $M_0$ at $z = 0$, takes a discrete time step $\Delta z$ backwards in time, splits the halo into smaller progenitors $M_0 \to M_1 + M_2$ (where the masses $M_1$ and $M_2$ are randomly drawn from a probabilistic merger rate), and iteratively applies the same procedure to all arising progenitors until $z_{\mathrm{obs}}$ is reached. The major splitting events that occur during these simulations represent cluster mergers and are highly important for the strong-lensing statistics. Using a simplistic model for the kinematics during cluster mergers, we determine the trajectories of all progenitors that arise during the merger tree simulation. In particular, we compute the angular positions of the progenitors on the observer's sky at redshift $z_{\mathrm{obs}}$. Applying this procedure to all initial haloes, we effectively populate the past null cone with spatially correlated haloes, some of which are actively merging.

In paper I, we emphasised that we employed two completely independent algorithms to compute the strong-lensing statistics: (1) a standard Monte Carlo approach to sample isolated haloes from a given mass function, and (2) our new method that incorporates cluster mergers. We did so in preparation of the following works, but also to independently cross-check the new merger algorithm. For self-consistency, the two methods had to be based on the original mass function of \citet{1974ApJ...187..425P}.

On the other hand, in paper II we found that the precise choice of the halo mass function has a significant impact on the GEV distribution of the largest Einstein radius. Because it is based on the overly idealistic theory of spherical collapse, the mass function of \citet{1974ApJ...187..425P} is known to underestimate the abundance of high-mass haloes, which in turn leads to lower expectation values for the largest Einstein radius in a given cosmological volume. Conversely, the mass function of \citet{ShethTormen1999} over-predicts the number of high-mass haloes, which shifts the GEV distribution to too high values. The \citet{TinkerMF} (Tinker) mass function is a compromise between these two extremes and is broadly accepted as a more accurate representation of the mass function determined from N-body simulations. Therefore, all calculations of paper III were based on the Tinker mass function. 

In this work, we aim to calculate the order statistics of the largest Einstein radii as realistically as possible, so that our results can be compared with real observations. It would therefore be desirable to replace the original merger algorithm by an alternative algorithm that generates halo catalogues consistent with the Tinker mass function.

\citet{2014MNRAS.440..193J} recently compared several different merger tree codes, finding that the algorithm proposed by \citet{2008MNRAS.383..557P} (PCH) reproduces the merger rates measured in N-body simulations most accurately. The basic idea of the PCH algorithm is quite simple: although merger algorithms based on the original extended Press-Schechter formalism are slightly inaccurate, they are certainly not completely unacceptable. Instead, they exhibit statistical properties (such as trends with mass and redshift) that agree well with those of merger trees constructed from N-body simulations. Parkinson and co-workers thus proposed to take the Press-Schechter merger rate as a starting point, and slightly perturb this quantity using an empirically motivated function. The probability for drawing a progenitor of mass $M_1$ from a parent halo of mass $M_0$ at redshift $z$ is then modified according to
\begin{equation}
\label{eq:sl4:PCH_merger_rate}
\frac{\mathrm{d}N}{\mathrm{d}M_1} \to G \left( M_0, M_1 \right) \, \frac{\mathrm{d}N}{\mathrm{d}M_1} \; ,
\end{equation}
with the empirical function
\begin{equation}
G \left( M_0, M_1 \right) = G_0 \, \left[ \frac{\sigma \left( M_1 \right)}{\sigma \left( M_0 \right)} \right]^{\gamma_1} \, \left[ \frac{ \delta \left( z \right)}{\sigma \left( M_0 \right)} \right]^{\gamma_1} \; ,
\end{equation}
where $\delta(z)$ is the linear density threshold for collapse and $\sigma \left( M \right)$ denotes the rms linear density fluctuation extrapolated to redshift $z = 0$ in spheres containing mass $M$ \citep{1993MNRAS.262..627L}. $G_0$, $\gamma_1$ and $\gamma_2$ are free empirical parameters of this ansatz, which PCH constrained by fitting their algorithm to merger trees extracted from the Millennium simulation \citep{2005Natur.435..629S}.

While testing our implementation of the PCH algorithm, we observed that the algorithm calibrated with the best-fitting parameters given by PCH still notably under-predicts the abundance of high-mass haloes ($\sim 10^{15} M_{\odot}/h$). More accurate results can be achieved with the parameters given by \citet{2008MNRAS.388.1361B}, who employed a slightly different fitting procedure. In this work, we thus used the values derived by \citet{2008MNRAS.388.1361B}: $G_0 = 0.605$, $\gamma_1 = 0.375$ and $\gamma_2 = -0.115$.

The performance of our algorithm for projecting merging galaxy clusters onto the past null cone is visualised in Fig.~\ref{fig:sl4:PCH_mass_function}. We show a random halo catalogue in the redshift range $0.5 < z < 1.0$ (full sky) that was generated with the PCH merger algorithm. As a reference, we also plot the theoretically expected values of the mass functions of Press-Schechter, Tinker, and Sheth-Tormen. The initial haloes at redshift $z = 0$ were drawn from the Tinker mass function.  Figure~\ref{fig:sl4:PCH_mass_function} confirms that the sampled catalogue agrees well with the expectations of the Tinker mass function at the considered redshift range. Yet, even with the parameters from \citet{2008MNRAS.388.1361B}, the generated mass function falls slightly short of high-mass haloes. The deviation is only moderate, however, and to fix this problem, one would have to run large N-body simulations containing sufficiently many clusters with masses $M > 10^{15} M_{\odot}/h$ and re-fit the PCH algorithm. Clearly, this would go beyond the scope of the present work. For the following sections, it is only important to remember that the generated halo catalogues lack a few high-mass objects, which renders our estimates of the order statistics more conservative. Moreover, if one is only interested in haloes with redshifts $z \gg 0$, the problem of too few high-mass haloes can be mitigated by starting the merger tree simulations at higher initial redshifts, $z_{\mathrm{ini}} > 0$.

We cannot use an independent approach here to sample single haloes (without mergers) because the PCH algorithm does not exactly reproduce the Tinker mass function. Whenever we compute the statistics of Einstein radii neglecting cluster mergers, we simply take the catalogues generated by the merger algorithm instead and ignore the spatial correlations of the haloes, that is, we simply treat all haloes as isolated objects.

\subsection{Suggestions to reduce the computing time}
\label{subsec:sl4:reduce_computing_time}

Increasing computational complexity is one main difficulty in estimating the order statistics of the largest Einstein radii including clusters mergers. As was shown in paper II, about $10^3$ mock universes have to be sampled for accurate fits of the GEV distribution. For each realisation, typically $\sim 10^6$ merger trees need to be simulated, and subsequently, their strong-lensing signals have to be evaluated. This whole procedure is computationally quite demanding.

However, there are several steps that greatly reduce the required wall-clock time. Since future studies will probably face similar problems, it might be useful to list some of these steps here:

\begin{itemize}

\item \textbf{Mass thresholds}: As shown previously (and again verified for this work), only massive haloes contribute to the order statistics of the largest Einstein radii. We therefore only sample haloes with masses $M > 10^{14} \, M_{\odot}/h$ at the initial redshift. While simulating their merger trees, we discard all progenitors whose mass falls below $5 \times 10^{12} \, M_{\odot}/h$, because subhaloes at least need to exceed $5\%$ of the main halo mass to notably perturb the Einstein radius (cf. paper I). 

As described above, for each initial halo, our merger routine projects a list of progenitors onto the past null cone. If the total mass of all progenitors falls below $10^{14} \, M_{\odot}/h$, we discard the system. Otherwise, we sort the progenitors in descending mass order, determine the Einstein radius of the most massive halo, and flag all progenitors that have been enclosed by the tangential critical curve. Next, we check if the total mass of the remaining unenclosed progenitors is larger than $10^{14} \, M_{\odot}/h$. If that is the case, we again sort them in descending mass order and compute the next Einstein radius. We repeat this procedure until no relevant unenclosed progenitors remain.

\item \textbf{Maximum separation}: When computing the tangential critical curve that surrounds a certain progenitor, we only take the neighbouring haloes into account (superposition of deflection angles) whose distance is smaller than the sum of both virial radii, $d \leq \left( r_{\mathrm{vir,1}} + r_{\mathrm{vir,2}} \right)$. Haloes that are farther away can safely be neglected because they do not notably perturb the lensing signal.

\item \textbf{Einstein radius threshold}: By experience, we know that the $n$-th largest Einstein radius of a mock realisation will certainly be larger than a lower threshold $\theta_{\mathrm{eff}}^{\mathrm{min}}$. Then, before computing the detailed shape of the tangential critical curve of a lens system, we first quickly estimate the size of the resulting Einstein radius using the following steps:

We loop over all relevant haloes in the field of view and treat them as isolated gravitational lenses. Because of their triaxial density profile, the projected surface mass density is ellipsoidal. We determine the major axis of the isodensity ellipses and compute the radius $\theta_{\mathrm{E}}^{\mathrm{est}}$ at which the tangential eigenvalue $\lambda_{\mathrm{t}}$ vanishes. A circle with radius $\theta_{\mathrm{E}}^{\mathrm{est}}$ encloses the full tangential critical curve of the halo and accordingly certainly over-estimates the real effective Einstein radius, $\theta_{\mathrm{eff}}^{\mathrm{est}} > \theta_{\mathrm{eff}}$.

If the sum of all estimated Einstein radii (plus some tolerance) is smaller than the threshold $\theta_{\mathrm{E,eff}}^{\mathrm{min}}$, we can safely discard the system and skip the expensive detailed lensing computations.

\item \textbf{Shoelace formula}: The effective Einstein radius is derived from the area enclosed by the tangential critical curve. This area $A$ can efficiently be computed using Stoke's theorem in two dimensions
\begin{equation}
A = \int_A \mathrm{d}A = \int_A \mathrm{rot} \, \vec{v} \; \mathrm{d}A = \int_{\partial A} \vec{v} \, \mathrm{d}s \; ,
\end{equation}
where $\vec{v}$ must be a vector field with $\mathrm{rot} \, \vec{v} = -\partial_y v_x + \partial_x v_y = 1$, such as $\vec{v} = 1/2 \, (-y, x)^{\mathrm{T}}$. The calculation of the area thus reduces to a simple line integral, which is far cheaper than standard connected-component labelling algorithms. Practically, the line integral is implemented by means of a loop over all boundary points, summing up the contributions using the so-called shoelace formula.

\item \textbf{Parallelisation}: The calculations can optimally (and trivially) be parallelised with OpenMP/MPI. The wall-clock time decreases inversely proportional to the number of CPUs used.

\end{itemize}

\section{Extreme value and order statistics with cluster mergers}
\label{sec:sl4:extreme_order_statistics_mergers}

\begin{figure*}
  \centering
  \includegraphics[width=0.45\textwidth]{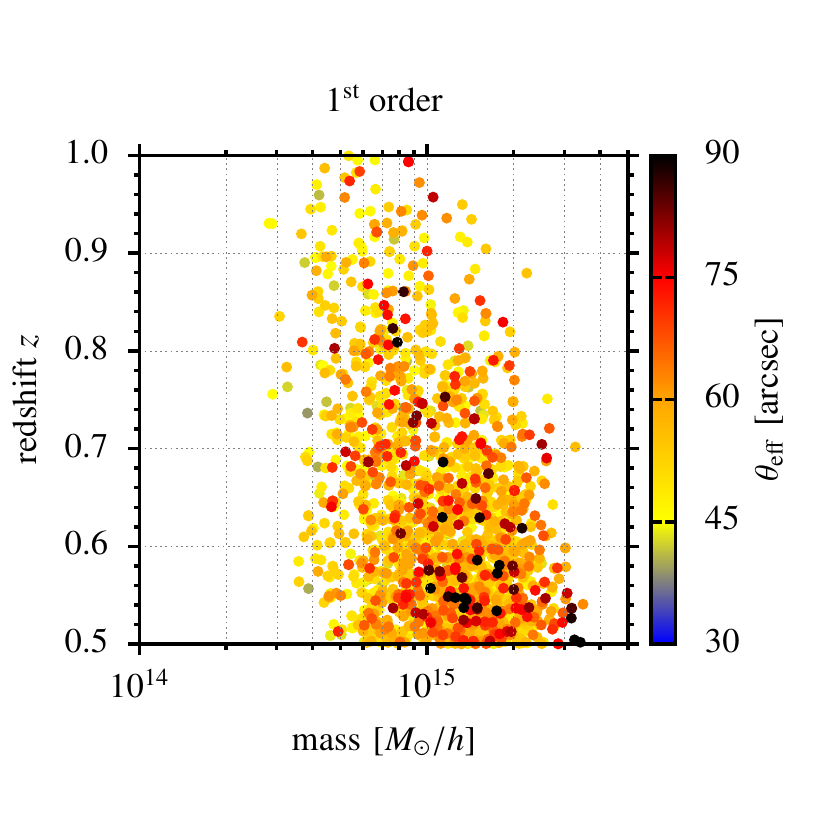}
  \includegraphics[width=0.45\textwidth]{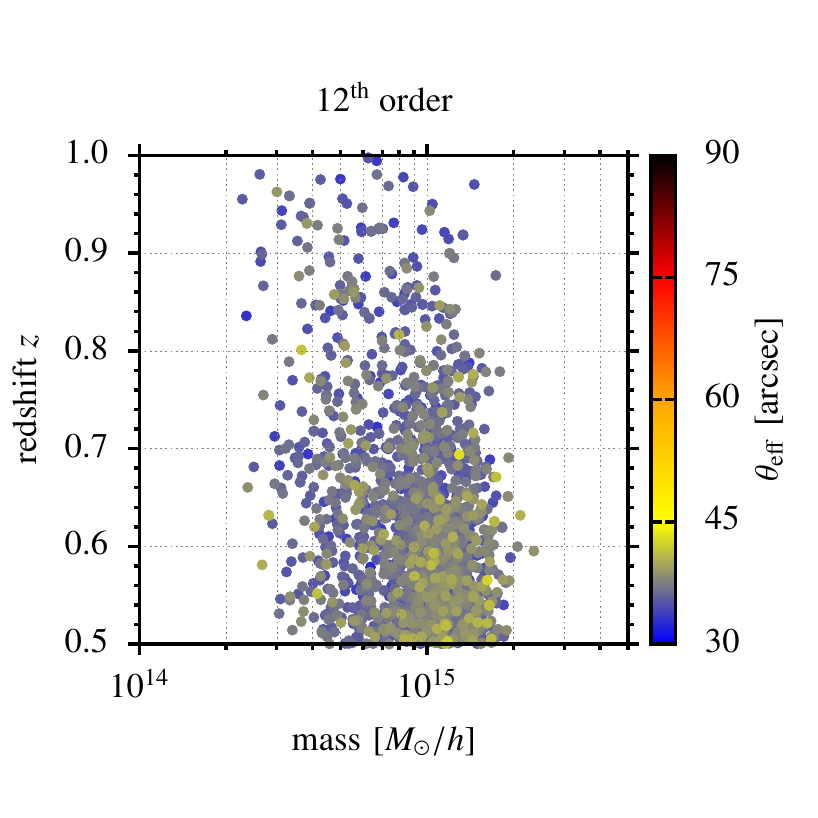}
  \caption[Distributions in mass and redshift of the lenses that produce the largest Einstein radii.]{Mass and redshift distributions of the gravitational lenses that produce the largest and the twelfth-largest Einstein radius. The distributions were extracted from 2000 mock realisations of Einstein radii in the redshift interval $0.5 \leq z \leq 1.0$ on the full sky. The calculations incorporated the impact of cluster mergers. The plotted mass denotes the total mass of all haloes enclosed by the corresponding tangential critical curves.}
  \label{fig:sl4:m_z_scatter_merger}
\end{figure*}

We now study the impact of cluster mergers on the order statistics of the largest Einstein radii in a certain cosmological volume. For comparison with our previous work (paper III), we exemplarily consider full-sky realisations of the cluster population in the redshift range $0.5 \leq z \leq 1.0$, which contains the twelve high-redshift ($z > 0.5$) MACS clusters analysed by \citet{2011MNRAS.410.1939Z}. In agreement with these authors, we assume a constant source redshift of $z_{\mathrm{s}} = 2.0$ throughout this work. We sampled 2000 mock realisations of the cluster population and collected the largest Einstein radius of each run. These data allowed us to accurately estimate the order statistics of the twelve largest Einstein radii.

Figure~\ref{fig:sl4:m_z_scatter_merger} shows the mass and redshift distributions of the clusters that produce the largest and the twelfth-largest Einstein radius. Because our analysis also includes merging clusters, it is important to stress that the mass referred to in Fig.~\ref{fig:sl4:m_z_scatter_merger} is the total mass of all haloes enclosed by the corresponding tangential critical curves. Generally, we note that cluster mergers do not significantly alter the mass and redshift distributions of the strongest gravitational lenses. The plots shown here are almost identical to those presented in paper III, which did not include the impact of cluster mergers. This is why we only show the plots of the first and the twelfth order here. Most importantly, however, Fig.~\ref{fig:sl4:m_z_scatter_merger} confirms that the lower mass threshold of $M_{\mathrm{min}} = 10^{14} \, M_{\odot}/h$ discussed in Sect.~\ref{subsec:sl4:reduce_computing_time} is well justified. 

\begin{figure}
  \centering
  \includegraphics[width=0.45\textwidth]{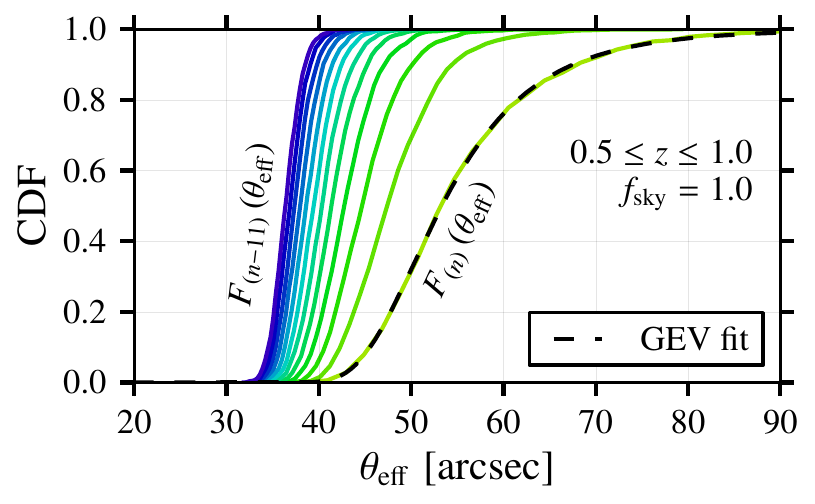}
  \caption[Cumulative distribution functions of the twelve largest Einstein radii, including the impact of cluster mergers.]{Cumulative distribution functions (CDF) of the twelve largest Einstein radii, including the impact of cluster mergers. The CDFs were extracted from 2000 mock realisations of Einstein radii in the redshift range $0.5 \leq z \leq 1.0$ on the full sky. The black dashed curve indicates the fit of the general extreme value (GEV) distribution to the CDF of the largest Einstein radius.}
  \label{fig:sl4:cdfs_orders_merger}
\end{figure}

Figure~\ref{fig:sl4:cdfs_orders_merger} shows the CDFs of the twelve largest Einstein radii extracted from the 2000 mock realisations sampled. While cluster mergers notably shift the CDFs to larger Einstein radii, the general characteristics discussed in paper III are conserved. The CDFs steepen with increasing order, implying that higher orders are, in principle, more constraining. The black dashed line indicates the excellent fit of the GEV to the CDF of the 2000 maxima sampled.

\begin{figure*}
  \centering
  \includegraphics[width=0.45\textwidth]{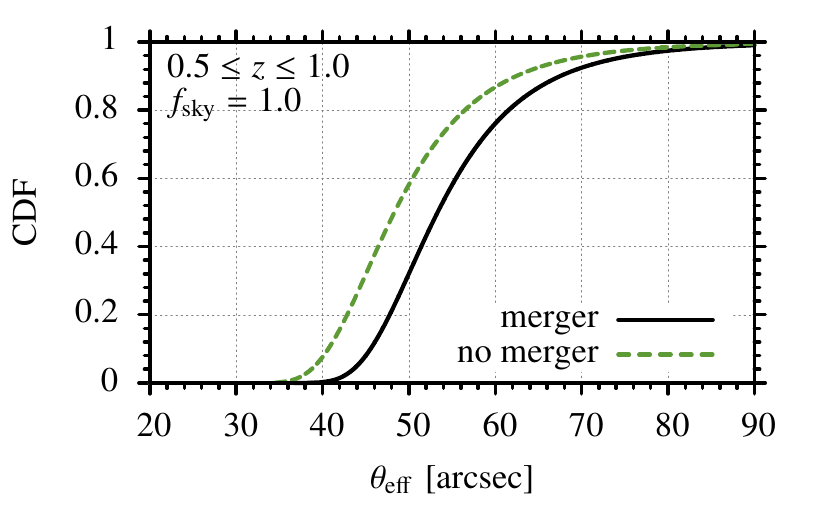}
  \includegraphics[width=0.45\textwidth]{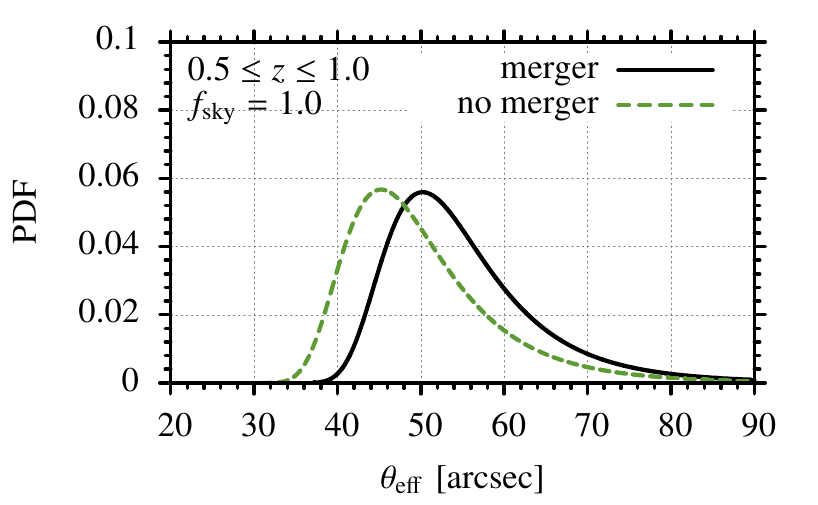}
  \caption[Impact of cluster mergers on the GEV of the largest Einstein radius.]{Impact of cluster mergers on the cumulative distribution function (CDF) and the probability density function (PDF) of the largest Einstein radius in the redshift range $0.5 \leq z \leq 1.0$ on the full sky. The general extreme value (GEV) distribution was fitted to the empiric CDF of maxima collected from 2000 mock realisations of the cosmological volume considered.}
  \label{fig:sl4:impact_merger_gev}
\end{figure*}

Next, we quantify the impact of cluster mergers on the GEV distribution of the largest Einstein radius. With mergers, the best-fitting parameters of the GEV distribution are given by $\left( \alpha, \beta, \gamma \right) = (50.85 \pm 0.02, 6.61 \pm 0.03, 0.100 \pm 0.005)$. Excluding mergers, the best-fitting parameters are $\left( \alpha, \beta, \gamma \right) = (45.88 \pm 0.02, 6.52 \pm 0.02, 0.105 \pm 0.004)$. The CDFs and the PDFs of these distributions are compared in Fig.~\ref{fig:sl4:impact_merger_gev}. In both cases, the shape parameters $\gamma$ are positive, indicating that the GEV distributions are bounded from below. Most importantly, mergers increase the location parameter $\alpha$ by $\sim 5 \arcsec$ (${\sim}11\%$), which means that the GEV distribution is significantly shifted to higher values. The mode of the GEV distribution -- which is the most likely value -- is given by
\begin{equation}
\label{eq:sl4:gev_mode}
 x_0=\alpha+\frac{\beta}{\gamma}\left[\left(1+\gamma\right)^{-\gamma}-1\right],
\end{equation}
and increases from $45.2 \arcsec$ without mergers to $50.2 \arcsec$ with mergers. The expectation value value of the GEV distribution reads
\begin{equation}
\label{eq:sl4:gev_ev}
\mathrm{E}_{\rm GEV}=\alpha-\frac{\beta}{\gamma}+\frac{\beta}{\gamma}\Gamma
\left(1-\gamma\right),
\end{equation}
where $\Gamma$ denotes the Gamma function. Again, the expectation value also increases from $50.4\arcsec$ to $55.4 \arcsec$. 

\begin{figure}
  \centering
  \includegraphics[width=0.48\textwidth]{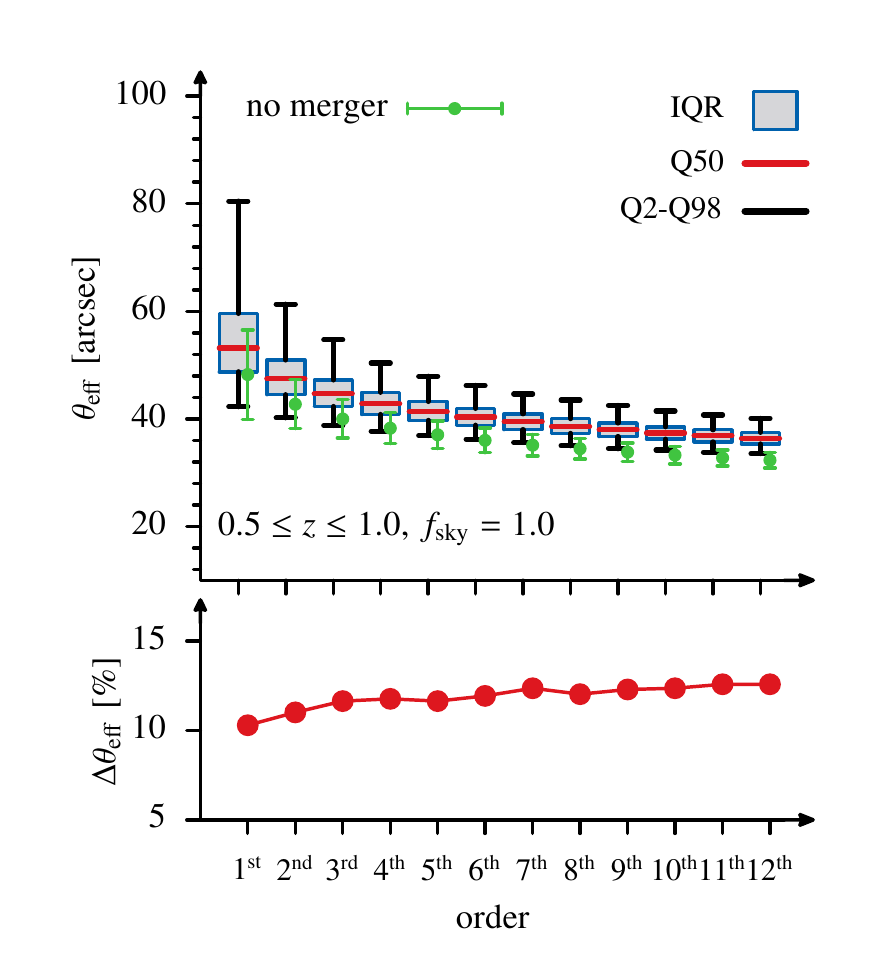}
  \caption[Box-and-whisker diagram for the order statistics of the twelve largest Einstein radii, including the impact of cluster mergers.]{Box-and-whisker diagram comparing the order statistics of the twelve largest Einstein radii including and excluding cluster mergers. For each order, in the upper plot the red lines indicate the median (Q50), the blue bordered grey boxes give the interquartile range (IQR), and the black whiskers mark the range between the 2 and 98 percentile (Q2, Q98) of the order statistics including the impact of cluster mergers. The green errors bars indicate the 68\% confidence intervals of the order statistics excluding mergers. The lower plot shows the percentage increase of the medians caused by mergers.}
  \label{fig:sl4:box_whisker_merger}
\end{figure}

The impact of cluster mergers on the order statistics of the twelve largest Einstein radii is summarised by the box-and-whisker diagram presented in Fig.~\ref{fig:sl4:cdfs_orders_merger}. We chose this representation to simplify the comparison with the results of paper III, but also because these diagrams compactly visualise important properties of statistical distributions. All medians of the twelve (six) highest orders are higher than $36 \arcsec$ ($43 \arcsec)$, indicating that the $\Lambda$CDM model predicts a dozen Einstein radii as large as $\sim 35 \arcsec - 55 \arcsec$ in the considered cosmological volume. If the 98 percentile were defined as exclusion criterion, one would need to observe approximately ten Einstein radii with $\theta_{\mathrm{eff}} \gtrsim 41 \arcsec$, five with $\theta_{\mathrm{eff}} \gtrsim 48 \arcsec$ or one large system with $\theta_{\mathrm{eff}} \gtrsim 80 \arcsec$, to claim tension with the expectations of the $\Lambda$CDM model. Current observational data certainly do not exceed these expectations \citep{2011MNRAS.410.1939Z}. Finally, the lower panel of Fig.~\ref{fig:sl4:box_whisker_merger} reveals that not only the expected value of the largest Einstein radius, but also those of the higher orders shown in the diagram, increases by $10-12\%$ because of the impact of cluster mergers.

In summary, the results of this section agree well with the findings of paper I: cluster mergers are significant for the statistics of the strongest gravitational lenses. As a rule of thumb, mergers increase the expected values of the largest Einstein radii by ${\sim}10\%$.

\section{Comparison with observational data}
\label{sec:sl4:comparison_observations}

\subsection{Galaxy cluster MACS J0717.5+3745}
\label{sec:sl4:obs_data_macs}

\begin{figure}
  \centering
  \includegraphics[width=0.45\textwidth]{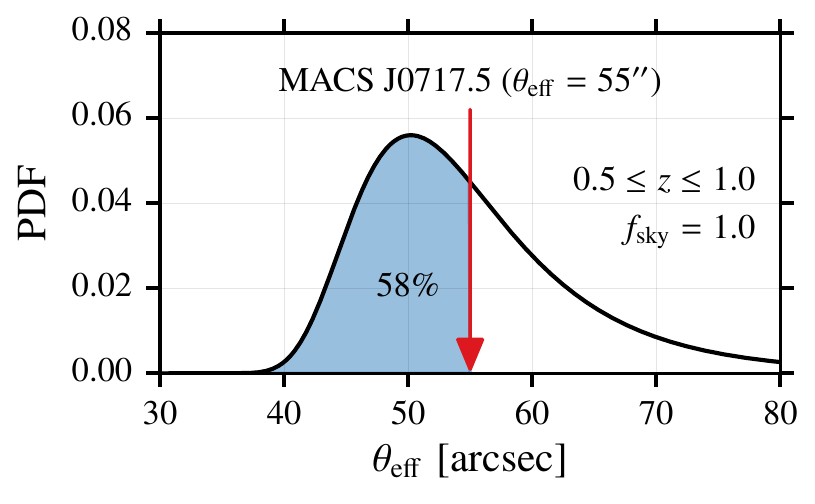}
  \caption[Comparison between the theoretical GEV distribution and the Einstein radius of MACS J0717.5+3745.]{Comparison between the theoretically expected general extreme value (GEV) distribution of the largest Einstein radius and the Einstein radius observed in the galaxy cluster MACS J0717.5+3745. The GEV distribution was fitted to the cumulative distribution of the largest Einstein radii extracted from 2000 mock realisations of the cluster population in the redshift range $0.5 \leq z \leq 1.0$ (full sky).}
  \label{fig:sl4:pdf_macs07175}
\end{figure}

As discussed in paper II, the X-ray luminous galaxy cluster MACS J0717.5+3745 is a remarkable system: it is extremely massive, actively merging, and exhibits the largest known Einstein radius \citep{2009ApJ...707L.102Z, 2011MNRAS.410.1939Z, 2012A&A...544A..71L}. The cluster was re-observed as part of the Cluster Lensing And Supernova survey with Hubble (CLASH; \citealt{2012ApJS..199...25P}), a 524-orbit Hubble Space Telescope (HST) multi-cycle treasury program. Thereupon, Zitrin and collaborators revised their initial mass model from 2009 as published in \citet{Medezinski2013}. The latest strong-lensing analysis of MACS J0717.5+3745 was conducted as part of the Hubble Frontier Fields program \footnote{http://www.stsci.edu/hst/campaigns/frontier-fields/}, yielding an Einstein radius of $\theta_{\mathrm{eff}} = \left( 55 \pm 6 \right)\arcsec$ for a source redshift $z_{\mathrm{s}} = 2.0$ (Zitrin et al., in preparation; private communication).

Although our case study in paper II already revealed that the Einstein radius of MACS J0717.5+3745 is not in tension with the expectations of the $\Lambda$CDM model, for completeness, we briefly update our results with the latest cosmological parameters from Planck \citep{2013arXiv1303.5076P}, and include the impact of cluster mergers. The cosmological volume analysed in the previous section was purposely chosen to contain the twelve MACS clusters discussed before. We can therefore use the best-fitting GEV parameters derived in Sect.~\ref{sec:sl4:extreme_order_statistics_mergers}.   

Figure~\ref{fig:sl4:pdf_macs07175} confirms that cluster mergers further confirm our main conclusion from paper II: the large Einstein radius of MACS J0717.5+3745 clearly does not exceed the theoretical expectations of the $\Lambda$CDM model. In contrast, the probability for observing a maximum Einstein radius even larger than $55 \arcsec$ amounts to ${\sim}42\%$. Neglecting cluster mergers, this probability decreases to ${\sim}24\%$ (as can easily be verified by evaluating the GEV distribution with the best-fitting parameters given in Sect.~\ref{sec:sl4:extreme_order_statistics_mergers}). The 98 percentile of the GEV distribution is located at $\theta_{\mathrm{eff}} = 82\arcsec$, indicating that extraordinarily strong lenses would have to be observed to claim disagreement with the $\Lambda$CDM model. Interestingly, when cluster mergers are included, the Einstein radius of MACS J0717.5+3745 (accidentally) coincides with the expectation value $\theta_{\mathrm{eff}} = 55 \arcsec$ of the theoretical GEV distribution (cf. Eq.~\eqref{eq:sl4:gev_ev}).

There is no need to revise the order statistics of the twelve MACS clusters discussed in paper III, because we already showed that most Einstein radii of the higher orders lie below the theoretically expected values, even if cluster mergers are excluded. As indicated by Fig.~\ref{fig:sl4:box_whisker_merger}, the inclusion of mergers would not add any new insight.

\subsection{SDSS clusters}
\label{sec:sl4:obs_data_sdss}

\begin{figure}
  \centering
  \includegraphics[width=0.48\textwidth]{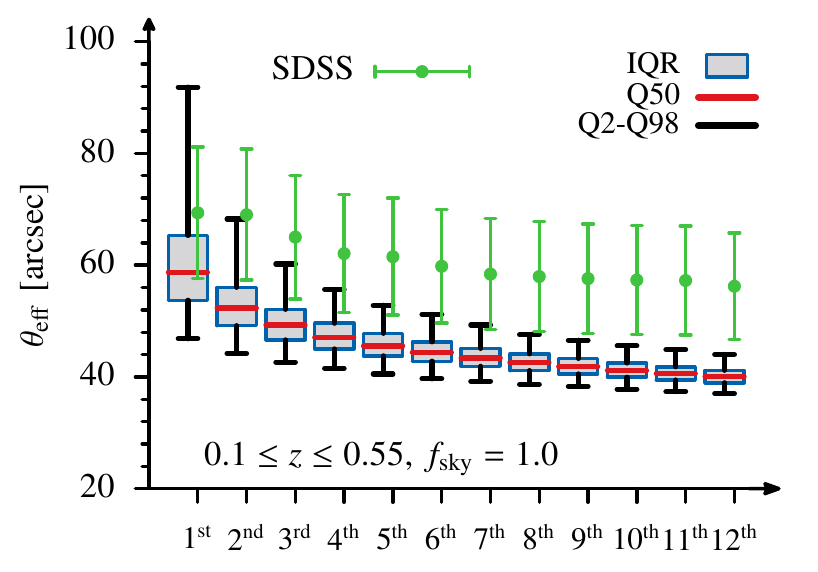}
  \caption[Box-and-whisker diagram comparing the twelve largest Einstein radii from the SDSS sample with the theoretically expected values.]{Box-and-whisker diagram comparing the theoretically expected order statistics of the twelve largest Einstein radii in the redshift range $0.1 \leq z \leq 0.55$ (full sky) and the largest Einstein radii in the SDSS sample analysed by \citet{2012MNRAS.423.2308Z}. For each order, the red lines indicate the median (Q50), the blue bordered grey boxes give the interquartile range (IQR), and the black whiskers mark the range between the 2 and 98 percentile (Q2, Q98) of the order statistics. The green error bars indicate the estimated 68\% confidence intervals of the Einstein radii in the SDSS sample.}
  \label{fig:sl4:box_whisker_sdss}
\end{figure}

\begin{table}
\centering
\caption{Comparison between the twelve largest Einstein radii in the SDSS sample analysed by \citet{2012MNRAS.423.2308Z} and the theoretically expected order statistics for the redshift range $0.1 \leq z \leq 0.55$ (full sky). The error bounds of columns two and three indicate the $1\sigma$ confidence intervals. Q98 denotes the 98 percentile of the theoretical order statistics.}
\begin{tabular}{c|c|cc}
  \hline\hline
   & SDSS & \multicolumn{2}{c}{theory} \\
  order & $\theta_{\mathrm{eff}} \, \left[ \mathrm{arcsec} \right]$ &  $\theta_{\mathrm{eff}} \, \left[ \mathrm{arcsec} \right]$ & Q98 [arcsec] \\
  \hline
  1\textsuperscript{st}  & $69\pm12$ & $59 \pm 9$ & 92 \\
  2\textsuperscript{nd}  & $69\pm12$ & $52 \pm 5$ & 68 \\
  3\textsuperscript{rd}  & $65\pm11$ & $49 \pm 4$ & 60 \\
  4\textsuperscript{th}  & $62\pm10$ & $47 \pm 3$ & 56 \\
  5\textsuperscript{th}  & $62\pm10$ & $46 \pm 3$ & 53 \\
  6\textsuperscript{th}  & $60\pm10$ & $44 \pm 3$ & 51 \\
  7\textsuperscript{th}  & $58\pm10$ & $43 \pm 2$ & 49 \\
  8\textsuperscript{th}  & $58\pm10$ & $43 \pm 2$ & 48 \\
  9\textsuperscript{th}  & $58\pm10$ & $42 \pm 2$ & 47 \\
  10\textsuperscript{th} & $57\pm10$ & $41 \pm 2$ & 46 \\
  11\textsuperscript{th} & $57\pm10$ & $41 \pm 2$ & 45 \\
  12\textsuperscript{th} & $56\pm10$ & $40 \pm 2$ & 44 \\  
  \hline
\end{tabular}
\label{tab:sl4:sdss_clusters}
\end{table}

\citet{2010ApJS..191..254H} applied a cluster-finding algorithm to the Sloan Digital Sky Survey (SDSS) Data Release 7 (DR7) data \citep{2009ApJS..182..543A} and assembled a large optical galaxy cluster catalogue consisting of over 55 000 rich clusters in the redshift range $0.1 \leq z \leq 0.55$. \citet{2012MNRAS.423.2308Z} proposed a new method for performing a simplistic strong-lensing analysis of these clusters in an automated way, and applied their technique to a subsample of 10 000 clusters to estimate their Einstein radii. 

The method proposed by \citet{2012MNRAS.423.2308Z} is based on the simple assumption that the light distribution observed in galaxy clusters generally traces their mass distribution well. The starting points are the red cluster member galaxies, which are assigned simple parametric mass profiles. The superposition of these individual profiles represents the galaxy component of the mass distribution. The dark matter component is constructed by smoothing the galaxies' distribution with a two-dimensional cubic spline interpolation. The sum of the two components serves as an indicator for the total projected matter density map of the cluster. The crucial point of the method is the calibration of the mass-to-light (M/L) ratio, which sets the normalisation of the projected surface mass density. \citet{2012MNRAS.423.2308Z} calibrated this M/L ratio using a subsample of ten well-studied galaxy clusters that were covered by both high-quality HST images and the SDSS, and assumed that this calibration is (approximately) valid for the entire SDSS sample. Using this procedure, Zitrin and co-workers automatically (and blindly) processed the 10 000 SDSS clusters, derived simple lens models from the photometry of the cluster member galaxies and estimated the corresponding Einstein radii for sources at redshift $z_{\mathrm{s}} = 2.0$. 

To compare their results with the theoretically expected order statistics of the largest Einstein radii, we sampled 2000 mock realisations of the redshift range $0.1 \leq z \leq 0.55$, assuming full sky coverage. Note that this choice was purposely too optimistic, because the SDSS DR7 only covered approximately one fourth of the full sky and, additionally, \citet{2012MNRAS.423.2308Z} analysed only ${\sim}20\%$ of the discovered clusters to reduce the required computing time. Our theoretical estimates should therefore be expected to exceed the distribution of Einstein radii determined for the subsample of SDSS clusters.

However, Fig.~\ref{fig:sl4:box_whisker_sdss} and Table~\ref{tab:sl4:sdss_clusters} reveal exactly the opposite trend. The mean estimated Einstein radii extracted from the SDSS sample significantly exceed the expected values, and the difference increases with higher orders. Starting with the second order, the mean estimated Einstein radii even exceed the 98 percentiles of the theoretical distributions. On the other hand, the errors of the SDSS Einstein radii are still large.  \citet{2012MNRAS.423.2308Z} suggested to assume an uncertainty of at least ${\sim}17\%$ for the $1\sigma$-boundary. From this point of view, all observed Einstein radii agree with the theoretically expected values within the $3\sigma$ confidence interval. However, we recall again that our theoretical estimates were computed for a substantially larger cosmological volume.

These results may imply that the observed gravitational lenses at low redshifts are somewhat stronger than theoretically expected, which would agree with the findings of \citet{2011MNRAS.418...54H}, for example. Given the current state of the observational data, however, this conclusion would certainly be premature. The uncertainties of the method used by \citet{2012MNRAS.423.2308Z} are still significant. Most importantly, the reliability of the assumed functional form for the M/L ratio needs to be verified and re-calibrated with a larger sample of well-studied gravitational lenses. Furthermore, Zitrin and co-workers mainly focused on the universal distribution of Einstein radii, instead of analysing the strongest lenses of their sample in detail (A. Zitrin, private communication). 

More meaningfully, we therefore conclude that the results of this section might indicate a discrepancy between theory and observations at low redshifts, and that it should certainly be interesting to carefully re-analyse the SDSS sample. Given the current state of the data and also the theoretical uncertainties influencing the modeling of the order statistics, there is, however, no reliable statistical evidence for claiming that these observations seriously challenge the predictions of the standard cosmological model.

\section{Conclusions}
\label{sec:sl4:conclusions}

This series of papers focused on the question whether or not the strongest observed gravitational lenses exceed the theoretical predictions of the standard cosmological model. In this last work, we combined all previously developed techniques and calculated the impact of cluster mergers on the order statistics of the largest Einstein radii. Moreover, we compared the theoretical results with observational data at different redshifts.

In the first part, we demonstrated that cluster mergers shift the GEV distribution of the largest Einstein radius to significantly higher values, typically increasing the location parameter and expected values by ${\sim}10\%$.  Furthermore, we showed that the order statistics of the $n$ largest Einstein radii are also shifted by a similar amount. This confirmed our findings of paper I, where we argued that cluster mergers are particularly important for the statistics of the strongest gravitational lenses.

In the second part, we compared the order statistics of the largest Einstein radii, including the impact of cluster mergers, with recent observational data. As already shown in papers II and III, at redshifts $z > 0.5$, we see no evidence for a tension between the strength of observed gravitational lenses and the theoretical predictions of the $\Lambda$CDM model. On the contrary, we actually find that the largest known Einstein radius at redshifts $z > 0.5$, which was observed in the galaxy cluster MACS J0717.5+3745 \citep{2009ApJ...707L.102Z, 2011MNRAS.410.1939Z, 2012A&A...544A..71L, Medezinski2013}, agrees excellently well with the theoretical expectation value.

At redshifts $z < 0.5$, the situation is a little more controversial. We compared the largest Einstein radii of the 10 000 SDSS clusters analysed by \citet{2012MNRAS.423.2308Z} to the theoretically expected order statistics, finding that the observed gravitational lenses in this redshift range appear to be stronger than expected. However, in this context, it is important to stress that the errors of the Einstein radii estimated for the SDSS sample are still large. Thus, given the premature state of current data, there is still no reliable statistical evidence for claiming disagreement with the $\Lambda$CDM model. Nevertheless, the SDSS sample might contain extraordinarily strong gravitational lenses, and it should certainly be interesting to carefully re-analyse these systems in future studies.

In this last paper of our series, it is probably worth to highlight some important conclusions that should be drawn from our studies.

Firstly, we explained that the theory of extreme value statistics -- and more generally, the order statistics of the $n$ largest Einstein radii -- provides a rigorous mathematical framework for discussing the likelihood of the largest observed Einstein radii, and also illustrated how exclusion criteria can be formulated. We developed fast, semi-analytic methods for computing theoretical forecasts, particularly emphasising that including cluster mergers is significant for accurate estimates of the order statistics. For this reason, future semi-analytic studies inevitably need to take the impact of cluster mergers into account before questioning the validity of the $\Lambda$CDM model. 

Secondly, we extensively discussed that the order statistics of the $n$ largest Einstein radii is strongly influenced by many model parameters. This property can both be interpreted as a strength and a weakness. It can be seen as a strength because it renders the order statistics a sensitive cosmological probe that reacts to changes of the cosmological parameters (e.g., $\Omega_{\mathrm{m}}$ and $\sigma_8$), details of structure formation (e.g., merger rates and halo mass function) and properties of triaxial dark matter haloes (e.g., concentration and axis ratios). On the other hand, it can be seen as a weakness because due to this sensitivity, the order statistics is currently still subject to many theoretical uncertainties. Given the current state of the theory, it is certainly difficult to formulate reliable exclusion criteria that seriously challenge the underlying cosmological model.

Finally -- and this is probably the most important result of this series of papers -- we can conclude that there is no statistical evidence at present for claiming that the largest observed Einstein radii exceed the theoretical expectations of the $\Lambda$CDM model.

\begin{acknowledgements}

We thank Adi Zitrin for valuable comments and for kindly providing the data set of 10 000 Einstein radii. We also thank Shaun Cole and Andrew J. Benson for helpful discussions. This work was supported in part by the German
\emph{Deut\-sche For\-schungs\-ge\-mein\-schaft, DFG\/} (DFG), and in part by contract research \textit{Internationale Spitzenforschung II-1} of the Baden-W\"{u}rttemberg 
Stiftung. Most simulations required for this work were performed on the bwGRiD cluster (http://www.bw-grid.de), member of the German D-Grid initiative, funded by the Ministry for Education and Research (Bundesministerium für Bildung und Forschung) and the Ministry for Science, Research and Arts Baden-Wuerttemberg (Ministerium für Wissenschaft, Forschung und Kunst Baden-Württemberg).

\end{acknowledgements}

\bibliographystyle{aa}
\bibliography{SL4_merger}

\begin{thebibliography}{27}
\expandafter\ifx\csname natexlab\endcsname\relax\def\natexlab#1{#1}\fi

\bibitem[{{Abazajian} {et~al.}(2009){Abazajian}, {Adelman-McCarthy},
  {Ag{\"u}eros}, {Allam}, {Allende Prieto}, {An}, {Anderson}, {Anderson},
  {Annis}, {Bahcall}, \& et~al.}]{2009ApJS..182..543A}
{Abazajian}, K.~N., {Adelman-McCarthy}, J.~K., {Ag{\"u}eros}, M.~A., {et~al.}
  2009, \apjs, 182, 543

\bibitem[{{Bartelmann}(2010)}]{2010CQGra..27w3001B}
{Bartelmann}, M. 2010, Classical and Quantum Gravity, 27, 233001

\bibitem[{{Benson}(2008)}]{2008MNRAS.388.1361B}
{Benson}, A.~J. 2008, \mnras, 388, 1361

\bibitem[{{Fisher} \& {Tippett}(1928)}]{1928PCPS...24..180F}
{Fisher}, R.~A. \& {Tippett}, L.~H.~C. 1928, Proceedings of the Cambridge
  Philosophical Society, 24, 180

\bibitem[{Gnedenko(1943)}]{gnedenko1943limited}
Gnedenko, B. 1943, Annals of Mathematics, 44, 423

\bibitem[{Gumbel(1958)}]{gumbel1958statistics}
Gumbel, E. 1958, Columbia Univ. press, New York

\bibitem[{{Hao} {et~al.}(2010){Hao}, {McKay}, {Koester}, {Rykoff}, {Rozo},
  {Annis}, {Wechsler}, {Evrard}, {Siegel}, {Becker}, {Busha}, {Gerdes},
  {Johnston}, \& {Sheldon}}]{2010ApJS..191..254H}
{Hao}, J., {McKay}, T.~A., {Koester}, B.~P., {et~al.} 2010, \apjs, 191, 254

\bibitem[{{Horesh} {et~al.}(2011){Horesh}, {Maoz}, {Hilbert}, \&
  {Bartelmann}}]{2011MNRAS.418...54H}
{Horesh}, A., {Maoz}, D., {Hilbert}, S., \& {Bartelmann}, M. 2011, \mnras, 418,
  54

\bibitem[{{Jiang} \& {van den Bosch}(2014)}]{2014MNRAS.440..193J}
{Jiang}, F. \& {van den Bosch}, F.~C. 2014, \mnras, 440, 193

\bibitem[{{Jing} \& {Suto}(2002)}]{2002ApJ...574..538J}
{Jing}, Y.~P. \& {Suto}, Y. 2002, \apj, 574, 538

\bibitem[{{Lacey} \& {Cole}(1993)}]{1993MNRAS.262..627L}
{Lacey}, C. \& {Cole}, S. 1993, \mnras, 262, 627

\bibitem[{{Limousin} {et~al.}(2012){Limousin}, {Ebeling}, {Richard},
  {Swinbank}, {Smith}, {Jauzac}, {Rodionov}, {Ma}, {Smail}, {Edge}, {Jullo}, \&
  {Kneib}}]{2012A&A...544A..71L}
{Limousin}, M., {Ebeling}, H., {Richard}, J., {et~al.} 2012, \aap, 544, A71

\bibitem[{{Medezinski} {et~al.}(2013){Medezinski}, {Umetsu}, {Nonino},
  {Merten}, {Zitrin}, {Broadhurst}, {Donahue}, {Sayers}, {Waizmann},
  {Koekemoer}, {Coe}, {Molino}, {Melchior}, {Mroczkowski}, {Czakon}, {Postman},
  {Meneghetti}, {Lemze}, {Ford}, {Grillo}, {Kelson}, {Bradley}, {Moustakas},
  {Bartelmann}, {Ben{\'{\i}}tez}, {Biviano}, {Bouwens}, {Golwala}, {Graves},
  {Infante}, {Jim{\'e}nez-Teja}, {Jouvel}, {Lahav}, {Moustakas}, {Ogaz},
  {Rosati}, {Seitz}, \& {Zheng}}]{Medezinski2013}
{Medezinski}, E., {Umetsu}, K., {Nonino}, M., {et~al.} 2013, \apj, 777, 43

\bibitem[{{Meneghetti} {et~al.}(2013){Meneghetti}, {Bartelmann}, {Dahle}, \&
  {Limousin}}]{2013SSRv..177...31M}
{Meneghetti}, M., {Bartelmann}, M., {Dahle}, H., \& {Limousin}, M. 2013, \ssr,
  177, 31

\bibitem[{{Parkinson} {et~al.}(2008){Parkinson}, {Cole}, \&
  {Helly}}]{2008MNRAS.383..557P}
{Parkinson}, H., {Cole}, S., \& {Helly}, J. 2008, \mnras, 383, 557

\bibitem[{{Planck Collaboration} {et~al.}(2013){Planck Collaboration}, {Ade},
  {Aghanim}, {Armitage-Caplan}, {Arnaud}, {Ashdown}, {Atrio-Barandela},
  {Aumont}, {Baccigalupi}, {Banday}, \& et~al.}]{2013arXiv1303.5076P}
{Planck Collaboration}, {Ade}, P.~A.~R., {Aghanim}, N., {et~al.} 2013,
  arXiv:1303.5076

\bibitem[{{Postman} {et~al.}(2012){Postman}, {Coe}, {Ben{\'{\i}}tez},
  {Bradley}, {Broadhurst}, {Donahue}, {Ford}, {Graur}, {Graves}, {Jouvel},
  {Koekemoer}, {Lemze}, {Medezinski}, {Molino}, {Moustakas}, {Ogaz}, {Riess},
  {Rodney}, {Rosati}, {Umetsu}, {Zheng}, {Zitrin}, {Bartelmann}, {Bouwens},
  {Czakon}, {Golwala}, {Host}, {Infante}, {Jha}, {Jimenez-Teja}, {Kelson},
  {Lahav}, {Lazkoz}, {Maoz}, {McCully}, {Melchior}, {Meneghetti}, {Merten},
  {Moustakas}, {Nonino}, {Patel}, {Reg{\"o}s}, {Sayers}, {Seitz}, \& {Van der
  Wel}}]{2012ApJS..199...25P}
{Postman}, M., {Coe}, D., {Ben{\'{\i}}tez}, N., {et~al.} 2012, \apjs, 199, 25

\bibitem[{{Press} \& {Schechter}(1974)}]{1974ApJ...187..425P}
{Press}, W.~H. \& {Schechter}, P. 1974, \apj, 187, 425

\bibitem[{{Redlich} {et~al.}(2012){Redlich}, {Bartelmann}, {Waizmann}, \&
  {Fedeli}}]{2012A&A...547A..66R}
{Redlich}, M., {Bartelmann}, M., {Waizmann}, J.-C., \& {Fedeli}, C. 2012, \aap,
  547, A66

\bibitem[{{Sheth} \& {Tormen}(1999)}]{ShethTormen1999}
{Sheth}, R.~K. \& {Tormen}, G. 1999, \mnras, 308, 119

\bibitem[{{Springel} {et~al.}(2005){Springel}, {White}, {Jenkins}, {Frenk},
  {Yoshida}, {Gao}, {Navarro}, {Thacker}, {Croton}, {Helly}, {Peacock}, {Cole},
  {Thomas}, {Couchman}, {Evrard}, {Colberg}, \& {Pearce}}]{2005Natur.435..629S}
{Springel}, V., {White}, S.~D.~M., {Jenkins}, A., {et~al.} 2005, \nat, 435, 629

\bibitem[{{Tinker} {et~al.}(2008){Tinker}, {Kravtsov}, {Klypin}, {Abazajian},
  {Warren}, {Yepes}, {Gottl{\"o}ber}, \& {Holz}}]{TinkerMF}
{Tinker}, J., {Kravtsov}, A.~V., {Klypin}, A., {et~al.} 2008, ApJ, 688, 709

\bibitem[{{Waizmann} {et~al.}(2012){Waizmann}, {Redlich}, \&
  {Bartelmann}}]{2012A&A...547A..67W}
{Waizmann}, J.-C., {Redlich}, M., \& {Bartelmann}, M. 2012, \aap, 547, A67

\bibitem[{{Waizmann} {et~al.}(2014){Waizmann}, {Redlich}, {Meneghetti}, \&
  {Bartelmann}}]{2014A&A...565A..28W}
{Waizmann}, J.-C., {Redlich}, M., {Meneghetti}, M., \& {Bartelmann}, M. 2014,
  \aap, 565, A28

\bibitem[{{Zitrin} {et~al.}(2011){Zitrin}, {Broadhurst}, {Barkana}, {Rephaeli},
  \& {Ben{\'{\i}}tez}}]{2011MNRAS.410.1939Z}
{Zitrin}, A., {Broadhurst}, T., {Barkana}, R., {Rephaeli}, Y., \&
  {Ben{\'{\i}}tez}, N. 2011, \mnras, 410, 1939

\bibitem[{{Zitrin} {et~al.}(2012){Zitrin}, {Broadhurst}, {Bartelmann},
  {Rephaeli}, {Oguri}, {Ben{\'{\i}}tez}, {Hao}, \&
  {Umetsu}}]{2012MNRAS.423.2308Z}
{Zitrin}, A., {Broadhurst}, T., {Bartelmann}, M., {et~al.} 2012, \mnras, 423,
  2308

\bibitem[{{Zitrin} {et~al.}(2009){Zitrin}, {Broadhurst}, {Rephaeli}, \&
  {Sadeh}}]{2009ApJ...707L.102Z}
{Zitrin}, A., {Broadhurst}, T., {Rephaeli}, Y., \& {Sadeh}, S. 2009, \apjl,
  707, L102

\end{thebibliography}

\end{document}